\documentclass[aps,prx,twocolumn,superscriptaddress]{revtex4-2}
\usepackage{amssymb,amsmath,amsfonts,bm}
\usepackage{graphicx}

\usepackage[usenames,dvipsnames]{color}
\newcommand{\fest}{f^\text{est}_\text{qubit}}
\newcommand{\ST}{S_\text{T}}
\newcommand{\SL}{S_\text{L}}

\begin{document}

\title{Coherence of a driven electron spin qubit actively decoupled from quasi-static noise}

\author{Takashi Nakajima}
\email[]{nakajima.physics@icloud.com}
\thanks{These authors contributed equally to this work.}

\author{Akito Noiri}
\thanks{These authors contributed equally to this work.}
\affiliation{Center for Emergent Matter Science, RIKEN, 2-1 Hirosawa, Wako-shi, Saitama 351-0198, Japan}

\author{Kento Kawasaki}
\affiliation{Department of Applied Physics, University of Tokyo, 7-3-1 Hongo, Bunkyo-ku, Tokyo 113-8656, Japan}

\author{Jun Yoneda}
\altaffiliation{School of Electrical Engineering and Telecommunications, University of New South Wales, Sydney, New South Wales 2052, Australia}
\affiliation{Center for Emergent Matter Science, RIKEN, 2-1 Hirosawa, Wako-shi, Saitama 351-0198, Japan}

\author{Peter Stano}
\affiliation{Center for Emergent Matter Science, RIKEN, 2-1 Hirosawa, Wako-shi, Saitama 351-0198, Japan}
\affiliation{Department of Applied Physics, University of Tokyo, 7-3-1 Hongo, Bunkyo-ku, Tokyo 113-8656, Japan}
\affiliation{Institute of Physics, Slovak Academy of Sciences, 845 11 Bratislava, Slovakia}

\author{Shinichi Amaha}
\affiliation{Center for Emergent Matter Science, RIKEN, 2-1 Hirosawa, Wako-shi, Saitama 351-0198, Japan}

\author{Tomohiro Otsuka}
\altaffiliation{JST, PRESTO, 4-1-8 Honcho, Kawaguchi, Saitama, 332-0012, Japan}
\altaffiliation{Research Institute of Electrical Communication, Tohoku University, 2-1-1 Katahira, Aoba-ku, Sendai 980-8577, Japan}
\affiliation{Center for Emergent Matter Science, RIKEN, 2-1 Hirosawa, Wako-shi, Saitama 351-0198, Japan}

\author{Kenta Takeda}
\affiliation{Center for Emergent Matter Science, RIKEN, 2-1 Hirosawa, Wako-shi, Saitama 351-0198, Japan}

\author{Matthieu R. Delbecq}
\altaffiliation{Laboratoire de Physique de l’Ecole normale sup\'{e}rieure, ENS, Universit\'{e} PSL, CNRS, Sorbonne Universit\'{e}, Universit\'{e} Paris-Diderot, Sorbonne Paris Cit\'{e}, Paris, France}
\affiliation{Center for Emergent Matter Science, RIKEN, 2-1 Hirosawa, Wako-shi, Saitama 351-0198, Japan}

\author{Giles Allison}
\affiliation{Center for Emergent Matter Science, RIKEN, 2-1 Hirosawa, Wako-shi, Saitama 351-0198, Japan}

\author{Arne Ludwig}
\author{Andreas D. Wieck}
\affiliation{Lehrstuhl f\"{u}r Angewandte Festk\"{o}rperphysik, Ruhr-Universit\"{a}t Bochum, D-44780 Bochum, Germany}

\author{Daniel Loss}
\affiliation{Center for Emergent Matter Science, RIKEN, 2-1 Hirosawa, Wako-shi, Saitama 351-0198, Japan}
\affiliation{Department of Physics, University of Basel, Klingelbergstrasse 82, CH-4056 Basel, Switzerland}

\author{Seigo Tarucha}
\email[]{tarucha@riken.jp}
\affiliation{Center for Emergent Matter Science, RIKEN, 2-1 Hirosawa, Wako-shi, Saitama 351-0198, Japan}

\date{\today}

\begin{abstract}
The coherence of electron spin qubits in semiconductor quantum dots suffers mostly from low-frequency noise. During the last decade, efforts have been devoted to mitigate such noise by material engineering, leading to substantial enhancement of the spin dephasing time for an idling qubit.
However, the role of the environmental noise during spin manipulation, which determines the control fidelity, is less understood.
We demonstrate an electron spin qubit whose coherence in the driven evolution is limited by high-frequency charge noise rather than the quasi-static noise inherent to any semiconductor device. We employed a feedback control technique to actively suppress the latter, demonstrating a $\pi$-flip gate fidelity as high as $99.04\pm 0.23\,\%$ in a gallium arsenide quantum dot.
We show that the driven-evolution coherence is limited by the longitudinal noise at the Rabi frequency, whose spectrum resembles the $1/f$ noise observed in isotopically purified silicon qubits.
\end{abstract}

\maketitle

\section{Introduction: Noise in Spin Qubits}
\begin{figure}
\includegraphics[width=0.4\textwidth]{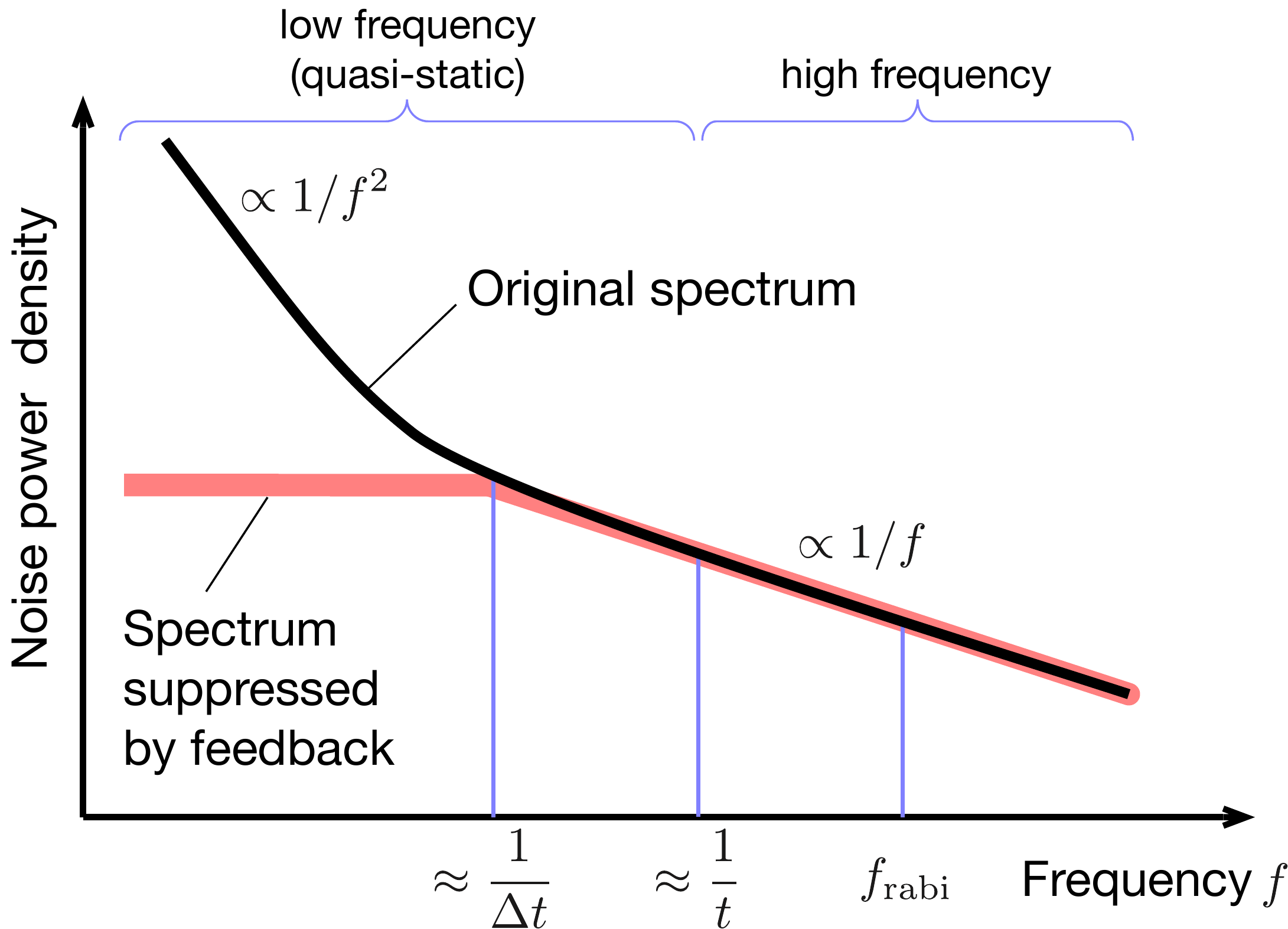}%
\caption{Example of noise power spectra for spin qubits with and without feedback.
A typical noise spectrum composed of $1/f^2$ and $1/f$ noise is shown in a log-log plot (black).
The feedback control acts like an active filter suppressing the low-frequency noise (red).
Shown on the bottom are relevant frequencies with $\Delta t$ the feedback latency, $t$ the qubit evolution time at which the coherence is evaluated, and $f_\text{rabi}$ the  Rabi frequency.
\label{fig:noise}}
\end{figure}

Since electrical manipulation of a single spin was demonstrated in semiconductor quantum dots\cite{Koppens2006}, enormous efforts have been devoted to improve spin coherence by controlling\cite{Foletti2009,Bluhm2010} or eliminating\cite{Veldhorst2014,Eng2015,Yoneda2017} nuclear spins, a magnetic noise source inherent to the host material\cite{Coish2004,Merkulov:2002ft,Khaetskii:2002jw}.
The progress is impressive: for example, dephasing times of $120\,\mu\text{s}$ in $^{28}$Si and $2\,\mu\text{s}$ in GaAs have been demonstrated\cite{Veldhorst2014,Shulman2014}. It is natural to expect that prolonging the spin coherence also improves the qubit control fidelity.
However, while the spin coherence is dominated by low-frequency (quasi-static) noise, control fidelity of a qubit is often impeded by noise at higher frequencies\cite{Martinis2003,Ithier2005,Lisenfeld2010,Yoshihara2014}.
The underlying relationship between the control fidelity and spin coherence remains elusive because there are different noise sources that could dominate in different frequency ranges, such as nuclear spin diffusion and charge fluctuators (see Fig.~\ref{fig:noise}). The former shows a $1/f^{\beta}$ spectrum with $3 > \beta > 1$ in GaAs\cite{Medford2012,Malinowski2017a} and possibly in natural Si devices\cite{Kawakami2016}, while the latter with $\beta \sim 1$ can dominate in $^{28}$Si devices\cite{Yoneda2017}.
In general, the dominant noise source depends on the material and structure of the quantum dot device as well as the frequency range of interest.
To understand the limits on the qubit control fidelity imposed by those different mechanisms, we build a feedback-controlled circuit which implements realtime Hamiltonian estimation\cite{Shulman2014}. It allows us to suppress the low-frequency noise\cite{Yang2019} and resolve the $1/f$ charge and nuclear spin noise at high frequencies.
We analyze how the low-frequency and high-frequency parts of the noise compete with each other and discuss the limitations of the high-fidelity control.

\section{Device and experimental setup}
\begin{figure}[h]
\includegraphics[width=0.8\textwidth]{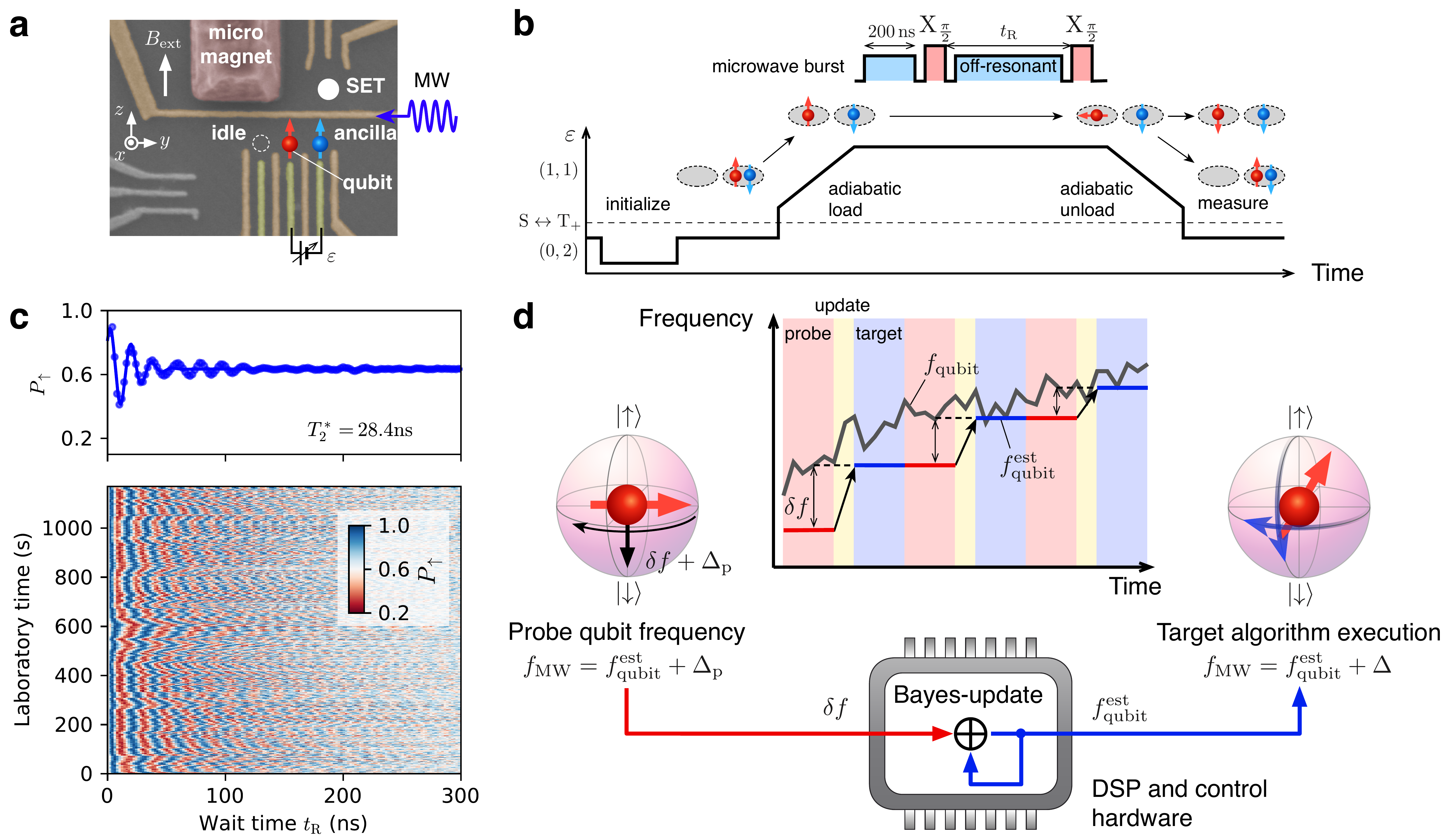}%
\caption{Ramsey meausurement and feedback-control scheme of an electron spin qubit.
(a) 
False-colored scanning electron micrograph image of the TQD device. An electron spin qubit in the middle QD (red arrow with a circle) is controlled by the EDSR where the spin is coupled to a microwave (MW) electric field via a stray magnetic field of the micromagnet deposited on the wafer surface\cite{Pioro-Ladriere:2008kx}. The right QD hosts an electron spin (blue arrow with a circle) used as a readout ancilla while the left QD hosts another electron which is unused and decoupled from the two spins. The energy detuning between the middle and the right QDs ($\varepsilon$) is gate-tunable and the QD electron occupancies are probed by a proximal single-electron transistor (SET)\cite{Barthel:2010fk}.
(b) 
Schematic of the Ramsey measurement. Two electrons (qubit and ancilla) are initialized to a doubly-occupied singlet state in the right QD and an up-spin qubit is prepared by adiabatically loading one of the electrons to the middle QD\cite{Noiri2016}. Two $\pi/2$ microwave bursts, separated by time $t_\text{R}$, are applied (before and during these, off-resonant microwave bursts are optionally applied). The ancilla-spin state is not affected by the microwave bursts. The final state is read out by unloading an up-spin (anti-parallel to the ancilla) state from the middle QD while a down-spin (parallel to the ancilla) state remains blocked.
(c) 
Up-spin probability $P_\uparrow$ as a function of $t_\text{R}$. The lower panel shows the Ramsey oscillations whose frequency varies with the laboratory time due to Overhauser field fluctuations. Each data point of $P_\uparrow$ is calculated from one hundred single-shot readout outcomes. The upper panel shows the trace obtained by averaging all the oscillations in the lower panel. The decay envelope gives the dephasing time of $T_{2}^{*}=28.4\,\text{ns}$, a value typical for electron spins in GaAs heterostructures.
(d) 
Schematic of the feedback control loop for a spin qubit. Data of a Ramsey oscillation as shown in (c) are processed in a digital signal processing (DSP) hardware with programmable logic (FPGA) to estimate the frequency detuning $\delta f = f_\text{qubit} - \fest$ between the current qubit frequency $f_\text{qubit}$ and its previous estimate $\fest$ (``probe'' step). The value of $\fest$ is updated to $\fest \mapsto \fest + \delta f$ (``update'' step), after which the target experiment follows (``target'' step). In the ideal case, the subsequent qubit algorithms can be executed with a microwave frequency $f_\text{MW}$ matching $f_\text{qubit}$ exactly (by choosing $\Delta=0$).
\label{fig:ramsey}}
\end{figure}

We use a triple quantum dot (TQD) device fabricated on a GaAs/AlGaAs heterostructure wafer. An electron is confined in each quantum dot (QD) by the electrostatic potentials induced by Ti/Au gate electrodes.
A Co micromagnet is placed on the surface and magnetized by a magnetic field of $B_\text{ext}=1.01\,\text{T}$ applied in the $z$-direction (see Fig.~\ref{fig:ramsey}a), creating inhomogeneous magnetic field over the QD array. The single electron spin qubit reported in this work is located in the middle QD and manipulated by the electric-dipole spin resonance (EDSR)\cite{Pioro-Ladriere:2008kx,Tokura:2006ir,Yoneda2014}. 
It is initialized and measured using the ancilla electron spin in the right QD\cite{Noiri2016}, see Fig.~\ref{fig:ramsey}b. An up-spin state of the qubit is prepared by initializing a doubly-occupied singlet ground state in the right QD and loading one of the electrons to the middle QD.
The voltage ramp is chosen to be adiabatic with respect to the inter-dot tunnel gap and the local magnetic field difference between the two dots but non-adiabatic with respect to the hyperfine gap.
The final state is read out by unloading an up-spin state to the right QD in the reverse process, while leaving a down-spin state blocked in the middle QD.
The experiment is conducted in a dilution refrigerator with an electron temperature of $120\,\text{mK}$.

\onecolumngrid

\twocolumngrid

We first perform a standard Rabi measurement\cite{Noiri2016} to roughly identify the Rabi frequency $f_\text{rabi}$ and the qubit resonance frequency $f_\text{qubit}=|g\mu_\text{B}B_\text{total}|/h$. Here $g$ is the electron $g$-factor, $\mu_\text{B}$ is the Bohr magneton, and $B_\text{total}$ is the sum of $B_\text{ext}$ and the $z$ components of the Overhauser field $B^\text{nuc}_{z}$ and the micromagnet stray field $B^\text{MM}_{z}$. After that, we measure Ramsey oscillations using two $\pi/2$ microwave bursts of duration $(4f_\text{rabi})^{-1}$ separated by a time interval $t_\text{R}$. The lower panel of Fig.~\ref{fig:ramsey}c shows the data gathered over $1200\,\text{s}$ with a fixed microwave frequency of $f_\text{MW}=5.55\,\text{GHz}$. The frequency of the measured oscillations fluctuates around a mean value $f_\text{MW}-f_\text{qubit} \approx 55\,\text{MHz}$. The fluctuations arise from changes of $B^\text{nuc}_z$ due to nuclear spin diffusion\cite{Reilly2008}, leading to inhomogeneous broadening of $f_\text{qubit}$. Averaging all the measured data results in damped oscillations shown in the upper panel of Fig.~\ref{fig:ramsey}c. Fitting it with a Gaussian envelope gives a spin dephasing time of $T_\text{2}^{*}=28.4\,\text{ns}$.

\section{The Feedback protocol}
To suppress this dephasing, rooted in slow fluctuations\cite{Coish2004} (quasi-static noise) of $f_\text{qubit}$, we employ a feedback-control scheme based on the realtime Hamiltonian estimation\cite{Giedke2006,Klauser2006}(see Fig.~\ref{fig:ramsey}d).
A similar technique was previously adopted for singlet-triplet qubits\cite{Shulman2014} to evaluate the stability of the idle qubit frequency, expressed by the dephasing time $T_\text{2}^{*}$, and its improvement upon noise estimation. Here we apply this technique to a single spin. The difference from a singlet-triplet qubit first of all requires changes in the protocol details, as given below. Second, the noise field couples to the spin differently, through its local value rather than its spatial gradient. However, the most important difference is that we focus here on the stability of the qubit being driven, rather than sitting idle. The rest of this section describes the details of the feedback protocol and its benchmarking, by examining how it boosts the dephasing time $T_\text{2}^{*}$. Readers interested solely on its benefits for the driven qubit stability can proceed directly to the next section.

The feedback scheme alternates the ``probe'' and ``target'' steps. In the former, the qubit frequency is probed by sampling $150$ up or down-spin outcomes of a Ramsey oscillation with $t_\text{R}=2,4,\cdots 300\,\text{ns}$ using $f_\text{MW}=\fest + \Delta_\text{p}$. Here, $\fest$ is the result of the qubit frequency estimation in the preceding probe step and $\Delta_\text{p}=50\,\text{MHz}$ is a fixed frequency offset inserted to ensure $f_\text{MW}>f_\text{qubit}$. With these settings, we use a Bayesian algorithm to estimate the instantaneous frequency detuning, $\delta f = f_\text{qubit} - \fest$. At the end of the probe step, the value of $\fest$ is updated to $\fest \mapsto \fest + \delta f$ and the microwave frequency is set to $f_\text{MW}=\fest+\Delta$. The subsequent ``target'' step begins after a short delay ($\sim$ms) to stabilize the signal output. The variable $\Delta$ is a controllable offset such that $\Delta = 0$ corresponds to the target algorithm executed with $f_\text{MW}$ equal to $f_\text{qubit}$.
By continuously looping the probe-target sequence, we can compensate low-frequency fluctuations of $f_\text{qubit}$ and remove their contribution to various qubit errors. For example, the dephasing time  $T_\text{2}^{*}$ is expected to be boosted by employing such compensation protocol.

\begin{figure}
\includegraphics[width=0.46\textwidth]{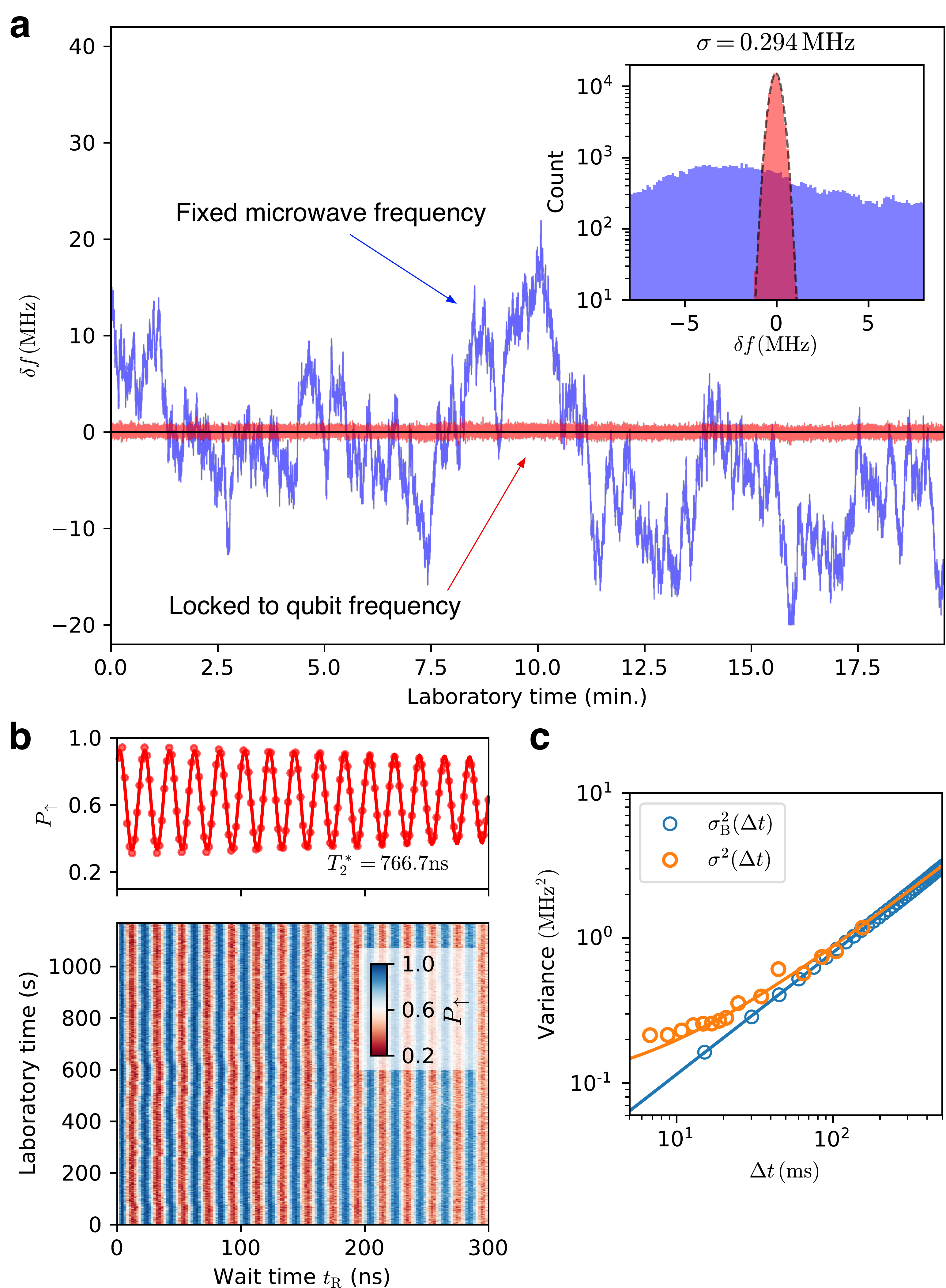}%
\caption{Suppressed dephasing of an electron spin qubit in a feedback-controlled rotating frame.
(a) 
Time dependence of the frequency detuning $\delta f = f_\text{qubit} - \fest$ extracted from Ramsey measurements. The blue trace is taken with a fixed $f_\text{MW}$ and the red trace is taken with a feedback-controlled $f_\text{MW}$.
The inset shows the histograms of $\delta f$ in the two cases. The histogram with the feedback control exhibits a normal distribution with a variance $\sigma^{2}=\left<(\delta f)^{2}\right>=(0.294\,\text{MHz})^2$ (black dashed curve).
(b) 
Ramsey oscillations as in Fig.~\ref{fig:ramsey}c but obtained with the feedback control ($\Delta = 50\,\text{MHz}$). The envelope of the oscillation in the upper panel is a Gaussian decay function drawn using dephasing time $T_{2}^{*}=1/(\pi\sqrt{2}\sigma)=766.7\,\text{ns}$.
(c) 
Variance of $\delta f$, $\sigma^{2} = \left<(\delta f)^{2}\right>$, as a function of the latency $\Delta t$ between the probe and target steps (orange circles) and that of the frequency correlator $\sigma_\text{B}^{2}$ in the laboratory frame (blue circles) as a function of the time difference $\Delta t$. A blue line is a fit to $\sigma_\text{B}^{2} = D\,\Delta t^\alpha$ and shows subdiffusive behaviour with the exponent $\alpha=0.84$ similar to a value found for singlet-triplet oscillations\cite{Delbecq2015}. The orange curve is a fit to $\sigma^{2} = D\,\Delta t^\alpha + (0.288\,\text{MHz})^{2}$.
\label{fig:feedback}}
\end{figure}

We now evaluate the performance of the feedback control by executing in the target step a Ramsey measurement similar to the one in the probe step ($\Delta = \Delta_\text{p} = 50\,\text{MHz}$). Figure~\ref{fig:feedback}a shows the values of $\delta f$ obtained from Ramsey measurements with feedback off ($f_\text{MW} = 5.55\,\text{GHz}$) and feedback on ($f_\text{MW}$ adjusted to $\fest$), respectively. The fluctuation of $\delta f$ is significantly suppressed by the feedback, exhibiting a Gaussian distribution with a variance $\sigma^{2}=(0.294\,\text{MHz})^{2}$ as shown in the inset. As a result, the Ramsey oscillation is substantially prolonged, as shown in Fig.~\ref{fig:feedback}b. Averaging the data with the overall acquisition time of $1200\,\text{s}$ gives a decay envelope well in line with the dephasing time expected from $\sigma$, being $T_{2}^{*}=1/(\pi\sqrt{2}\sigma)=766.7\,\text{ns}$.

The gain of $T_{2}^{*}$ is therefore directly associated with the achievable value of $\sigma$, which in turn is limited by the resolution of the Bayesian estimation. To demonstrate this, we plot $\sigma^{2}$ in Fig.~\ref{fig:feedback}c as a function of the feedback latency $\Delta t$, defined as the time interval between the probe and target steps plus the average time spent in each step (see Appendix A). For large latency, $\sigma^{2}$ approaches the variance of the qubit frequency correlator in the laboratory frame, $\sigma_\text{B}^{2}(\Delta t) \equiv \left<(f_\text{qubit}(t+\Delta t) - f_\text{qubit}(t))^{2}\right>$. It implies that in this regime $\sigma^{2}$ is dominated\cite{Delbecq2015} by the Overhauser field fluctuations during $\Delta t$. For small latency, $\sigma^2$ converges to $(0.288\,\text{MHz})^2$. This value is comparable to the bin width of the frequency discretization ($0.25\,\text{MHz}$) used in the Bayesian estimation algorithm performed by the feedback hardware. We believe that the variance $\sigma^{2}$ could be further decreased by using a smaller bin width as $\sigma_\text{B}^{2}$ continues to decrease with $\Delta t$ within the measured range, although we were not able to do so due to hardware limitations.
The value of $\sigma^{2}$, and thereby that of $T_{2}^{*}$, can be controlled by $\Delta t$, allowing for studying the effects of noise in different frequency ranges.

\onecolumngrid

\begin{figure}[h]
\includegraphics[width=0.7\textwidth]{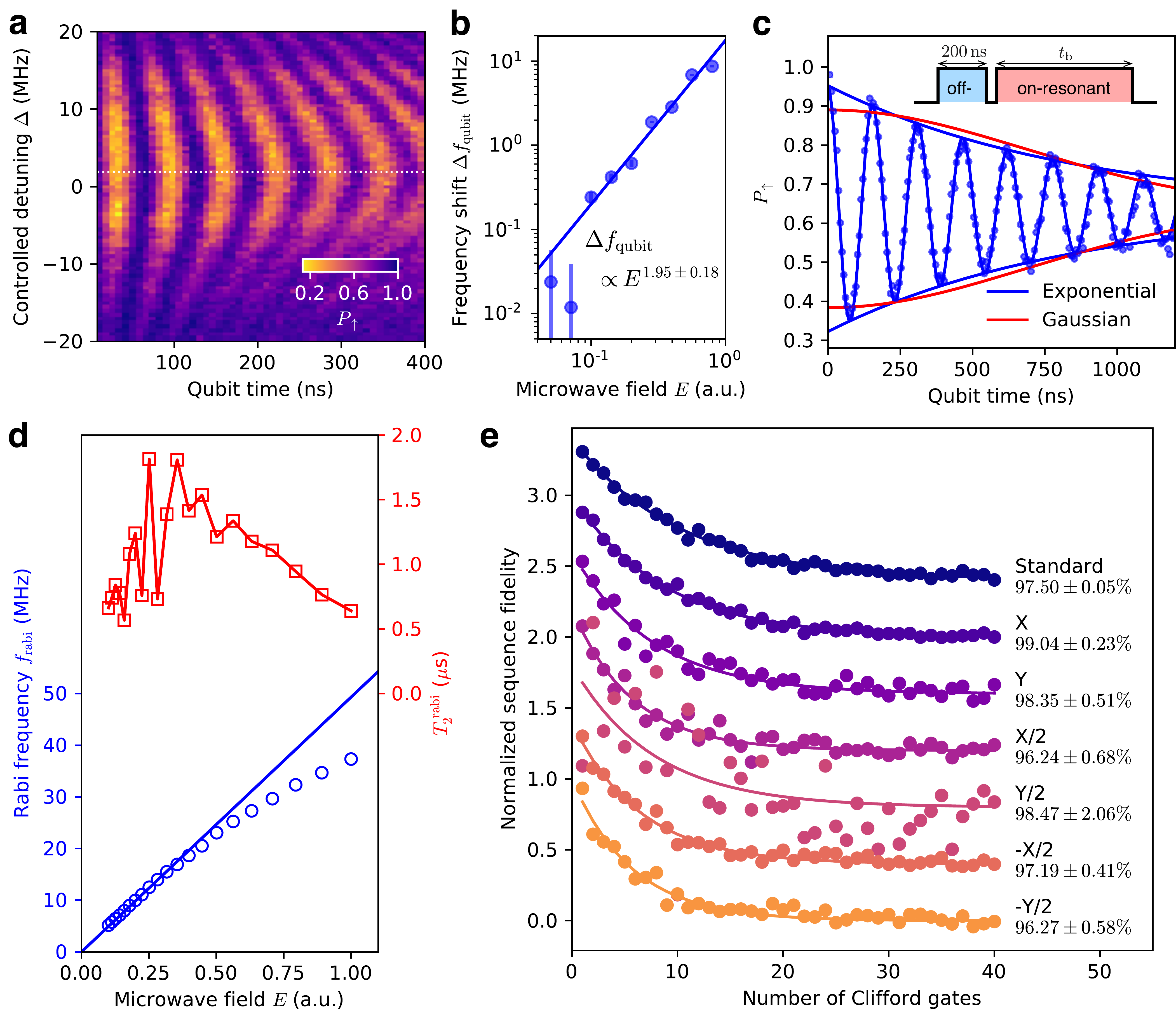}%
\caption{Coherent control and benchmarking of a single-electron spin with the feedback.
(a) 
Typical Rabi oscillations versus the offset $\Delta$ obtained in the feedback-controlled rotating frame. A horizontal white dashed line indicates the microwave-induced shift of the qubit frequency, $\Delta f_\text{qubit}$.
(b) 
Microwave-amplitude dependence of the qubit frequency shift. A blue line is a fit to the data showing that the amplitude dependence is quadratic ($\propto E^{1.95\pm 0.18}$).
(c) 
Rabi oscillations obtained at zero detuning upon compensating for the induced shift $\Delta f_\text{qubit}$. Namely, an off-resonant microwave burst with the same amplitude as the one used for the Rabi measurement is applied before and during the Ramsey measurement (see Fig.~\ref{fig:ramsey}c). A similar off-resonant burst is also applied for $200\,\text{ns}$ before the Rabi drive to stabilize the value of the frequency shift.
(d) 
Microwave amplitude dependence of the Rabi frequency (blue circles) and the Rabi decay time (red squares with lines). A blue line is a linear fit to the low-amplitude data of the Rabi frequency.
(e) 
Normalized sequence fidelities for standard (top) and interleaved (others) randomized benchmarking. Traces are offset by $0.4$ for clarity. The standard sequence shows an exponential decay with an average single-gate fidelity of $97.50\pm 0.05\%$. Interleaved sequences are annotated with corresponding single-qubit gates and extracted fidelities.
\label{fig:control}}
\end{figure}

\twocolumngrid

\section{Improvements of the qubit control}
We now turn to benchmarking of the qubit-control fidelity with the boosted dephasing time.
Figure~\ref{fig:control}a shows Rabi oscillations of the single-spin qubit measured with a varied frequency offset $\Delta$. We observe a clear chevron pattern with the symmetry axis offset from $\Delta=0$. This implies that the qubit frequency $f_\text{qubit}$ is shifted by a finite ac electric field $E$ of the driving microwave burst, while the frequency probed in a Ramsey measurement corresponds to that of $E=0$. The magnitude of this shift, $\Delta f_\text{qubit}(E,t) = f_\text{qubit}(E,t) - f_\text{qubit}(0,t)$, increases quadratically with $E$, as shown in Fig.~\ref{fig:control}b. Similar frequency shifts are observed in silicon devices with micromagnets, attributed to a spatial displacement of the electron wavefunction\cite{Watson2017,Takeda2018}. This effect is detrimental to our feedback protocol because $\Delta f_\text{qubit}(E,t)$ may vary with time $t$ due to the spatial dependence of the Overhauser field. In order to obtain $\fest=f_\text{qubit}(E,t)$ directly in the probe step, we apply an off-resonant microwave burst at $f_\text{off}=5.4\,\text{GHz}$ [$f_\text{qubit}-f_\text{off}>200\,\text{MHz}$] during the interval of $t_\text{R}$ which induces nominally the same displacement and the same $\Delta f_\text{qubit}(E,t)$ as for the target step. In addition, a $200\,\text{ns}$-long off-resonant pre-burst is applied to stabilize a transient component of the microwave-induced frequency shift\cite{Takeda2018}. The frequency is switched between $f_\text{off}$ and $f_\text{MW}$ within $1\,\text{ns}$ using a high-speed microwave switch (see Supplemental Material). We use this modified Ramsey sequence in the probe step and focus on the target step performed at zero detuning [$\Delta = 0,\, \fest = f_\text{qubit}(E,t)$] in all measurements described below.

Figure~\ref{fig:control}c shows a typical Rabi oscillation at zero detuning. It shows an exponential decay, which is common for silicon QDs\cite{Takeda2016,Yoneda2017} but atypical for GaAs\cite{Yoneda2014,Nichol2016}. We extract the Rabi frequency $f_\text{rabi}$ and the decay time of the driven oscillation $T_{2}^\text{rabi}$ from fitting and plot their dependence on the driving field amplitude in Fig.~\ref{fig:control}d.
For a given field amplitude, we find that both $f_\text{rabi}$ and $T_{2}^\text{rabi}$ are influenced by inter-dot detuning energy that modulates spin-electric-coupling (SEC) strength (see Appendix \ref{sec:extraction_of_PSD}).
We therefore optimize the detuning energy for the highest quality factor, defined as the number of typical qubit operations available within the Rabi decay time. We reach a quality factor of $Q=2f_\text{rabi}T_{2}^\text{rabi}=85\pm 8$ (see Fig.~\ref{fig:ramsey_SEC}b of Appendix \ref{sec:extraction_of_PSD}), comparable to natural silicon quantum dots\cite{Kawakami2016,Takeda2016}. The value predicts the fidelity of a $\pi$-gate of $e^{-1/Q}=98.8\pm 0.1\,\%$. We tested this prediction using randomized benchmarking\cite{Muhonen2015} and find an $\text{X}_\pi$ gate fidelity of $99.04\pm 0.23\,\%$ (see Fig.~\ref{fig:control}e), close to the $Q$-factor limited value. This is the highest fidelity for single-spin qubits in GaAs reported so far. We notice however that the average single-gate fidelity is $97.50\pm 0.05\,\%$, most likely limited by systematic errors in the other gates (unitary errors) due to the microwave setup in the present study (see Supplemental Material). This issue would be readily resolved by integrating an established technique of IQ modulation\cite{Takeda2016,Yang2019} with the FPGA in the microwave generation setup. The $\text{X}_\pi$ gate is least affected by microwave imperfections as we calibrate the control pulse line primarily for this gate.

\section{Discussion: limits on the spin-qubit control fidelity}
What is the physical mechanism limiting the Rabi decay time and the ultimate control fidelity of a single-spin qubit in this system? One obvious candidate is the residual inhomogeneity of $f_\text{qubit}$ or, in other words, the quasi-static noise $\delta f$.
However, this contribution should lead to a power-law envelope\cite{Koppens2007,Cucchietti2005} of the Rabi decay $[1+(2\pi\sigma^2 t/f_\text{rabi})^2]^{-1/4}$, as opposed to the exponential one seen in Fig.~\ref{fig:control}c. For $f_\text{rabi}\gg \sigma$, the initial decay could be approximated by a Gaussian envelope with\cite{Yoneda2014} $T_{2}^{\text{rabi}^\prime}=f_\text{rabi}/(\pi\sigma^{2})$, leading to $T_{2}^{\text{rabi}^\prime}=74\,\mu\text{s}$ with $f_\text{rabi}=20\,\text{MHz}$. This value is an order of magnitude larger than the measured decay time. Also, assuming that unitary errors are removed\cite{Yang2019}, the qubit control fidelity as high as $\exp\left[-(2f_\text{rabi} T_{2}^{\text{rabi}^\prime})^{-2}\right]>99.9999\,\%$ could be reached. We conclude that such quasi-static noise is therefore not the main limiting factor of the ultimate qubit control fidelity.

We consider three other noise sources as possibly relevant to the Rabi decay\cite{Yan2013,Yoshihara2014}: the quasi-static noise in $f_\text{rabi}$, the transverse noise at the electron Larmor frequency leading to spin relaxation, and the longitudinal noise in $f_\text{qubit}$ at the Rabi frequency. The quasi-static noise in $f_\text{rabi}$ could be caused by fluctuations of the microwave driving amplitude or SEC. However, this mechanism would also lead to a Gaussian decay envelope with $T_{2}^\text{rabi}\propto f_\text{rabi}^{-1}$, inconsistent with Fig.~\ref{fig:control}c. The spin relaxation is also unlikely, at least in the range of $f_\text{rabi}<20\,\text{MHz}$, because it cannot explain the increase of $T_{2}^\text{rabi}$ with $f_\text{rabi}$. Therefore, we conclude that $T_{2}^\text{rabi}$ is most likely dominated by the high-frequency (on the order of $f_\text{rabi}$) longitudinal noise in $f_\text{qubit}$, which inherently leads to an exponential Rabi decay.

\begin{figure}
\includegraphics[width=0.9\textwidth]{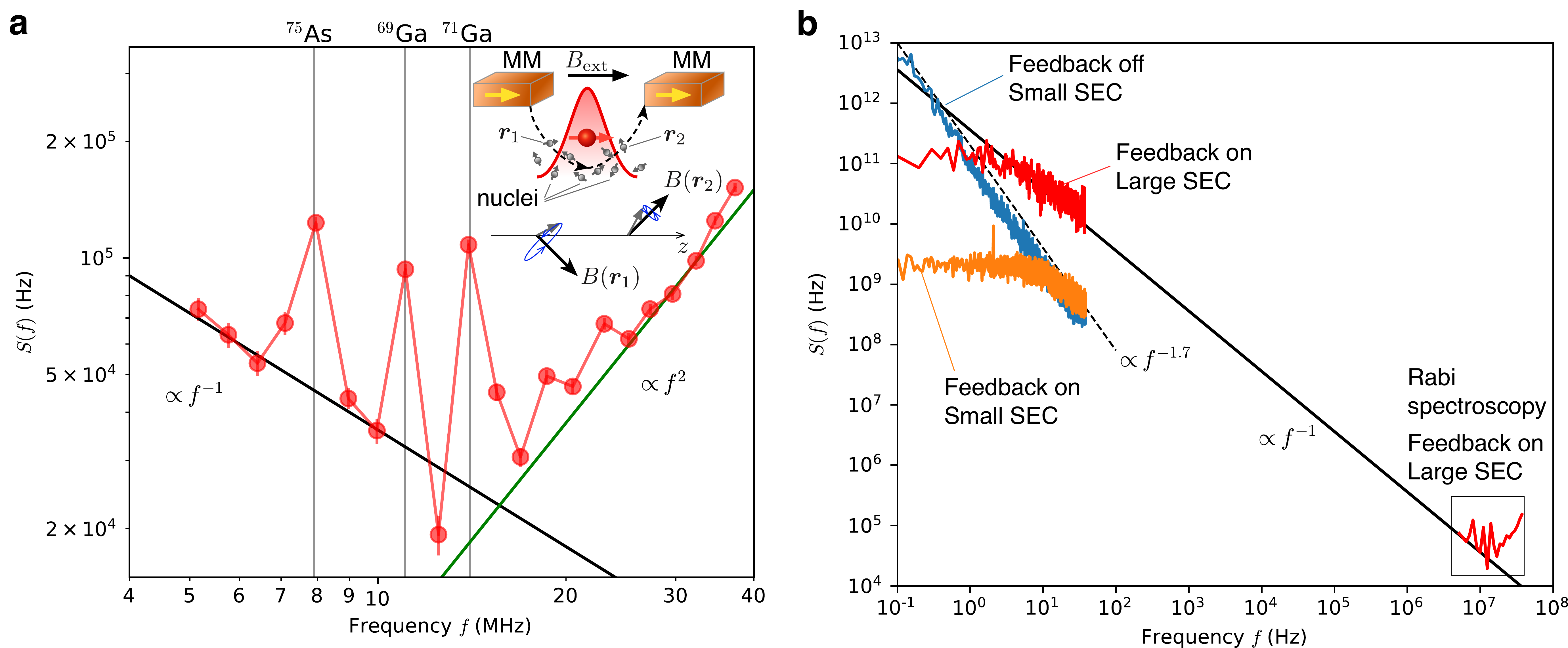}%
\caption{Rabi noise spectroscopy.
(a) 
Power spectral density $S(f)$ of the longitudinal noise in $f_\text{qubit}$ (red circles) extracted from the data in Fig.~\ref{fig:control}d. Vertical grey lines show Larmor precession frequencies for three nuclear species, $^{75}$As, $^{69}$Ga and $^{71}$Ga calculated with the micromagnet-induced field component of $B^\text{MM}_z=70\,\text{mT}$. Black and green lines are guides to the eye for $f^{-1}$ and $f^{2}$ dependence, respectively.
The inset illustrates electron-nuclear spin coupling in an inhomogeneous magnetic field. Each nuclear spin is randomly oriented and precesses around a local magnetic field vector $\bm{B}(\bm{r}_k)$, leading to an oscillatory Overhauser field in the $z$ direction.
(b) 
Comparison of $S(f)$ in (a) and those extracted from Ramsey measurements. The power spectral density of $\delta f$ in Ramsey measurements is calculated by the fast Fourier transform of $\delta f(t)$. The spectral density taken with small SEC and feedback off (blue) shows the $f^{-1.7}$ dependence (black dashed line) similar to those observed for nuclear spin diffusion noise\cite{Medford2012,Malinowski2017a}. The spectral density is significantly suppressed with feedback on (orange) down to a level determined by the precision of the feedback control. The flat noise spectrum suggests that the residual low-frequency noise is uncorrelated. A peak at $2\,\text{Hz}$ is due to the vibration of the dilution refrigerator. The power spectral density increases as SEC is increased (red curve in the top left), and it follows the $f^{-1}$ dependence (black solid line) in line with $S(f)$ extracted from the Rabi spectroscopy in (a) (red curve in the bottom right).
\label{fig:spectroscopy}}
\end{figure}

When the high-frequency noise in $f_\text{qubit}$ dominates, one can relate\cite{Yan2013,Yoshihara2014} the noise power spectral density $S(f)$ at the Rabi frequency to the exponential-decay rate of the Rabi oscillations [see Appendix \ref{sec:extraction_of_PSD}].
The power spectral density $S(f)$ extracted in this way is plotted in Fig.~\ref{fig:spectroscopy}a.
For $f>20\,\text{MHz}$, it grows rapidly with $f$. It could be due to thermal noise caused by microwave-induced heating\cite{Takeda2018}, with the consistent scaling $P\propto E^{2}\propto f_\text{rabi}^{2}$. On the other hand, we cannot exclude a Rabi decay through spin relaxation in this range, possibly caused by, for example, electron exchange with reservoirs due to photon-assisted tunneling\cite{Fujisawa1997,Oosterkamp:1998tm}. Since we could not extract more information of the intrinsic noise density, we do not discuss this range further.

We turn to the other frequency range, $f<20\,\text{MHz}$. Here, $S(f)$ shows three prominent peaks at nuclear Larmor precession frequencies of $^{75}$As, $^{69}$Ga and $^{71}$Ga. It clearly suggests that such high-frequency noise sources indeed influence the Rabi decay of the spin qubit. The hyperfine coupling between a single electron spin $\hat{\bm{S}}$ and nuclear spins $\hat{\bm{I}}_k$ is given by $H_\text{hf}=\sum_k A_k \hat{\bm{S}}\cdot\hat{\bm{I}}_k$, where $A_k$ is a coupling constant dependent on each nuclear site $\bm{r}_k$ indexed by $k$. This results in an Overhauser field component parallel to the external field, being $B^\text{nuc}_{z}\propto \sum_k A_k \hat{I}_k^z$. When nuclear spins precess around the $z$ axis, $B^\text{nuc}_{z}$ is constant and there is no noise at the Larmor frequencies. In the present device, however, the stray magnetic field from the micromagnet induces field inhomogeneity, making each nuclear spin at $\bm{r}_k$ precess around a local magnetic field vector $\bm{B}(\bm{r}_k)$ slightly off the $z$ axis (see the inset of Fig.~\ref{fig:spectroscopy}a). The inhomogeneity of the nuclear spin polarization leads to small but finite residual oscillations of $B^\text{nuc}_{z}$ at the nuclear precession frequencies.

\onecolumngrid

\twocolumngrid

Apart from the three spectral peaks, we find that $S(f)$ follows $f^{-1}$ dependence at $f<20\,\text{MHz}$. It suggests that the $f^{-1}$ noise background in the range of tens of MHz is the dominant limiting factor of the qubit control fidelity. The $f^{-\beta}$ dependence with $\beta=1$ differs from $\beta \sim 1.7$ for the nuclear spin diffusion noise\cite{Medford2012,Malinowski2017a} at low frequencies (see Fig.~\ref{fig:spectroscopy}b) extracted from the data in Figs.~\ref{fig:ramsey} and \ref{fig:feedback} (small-SEC regime). In addition, we confirmed that the amplitude of $S(f)$ is larger for a larger SEC (See Fig.~\ref{fig:sec_comparison} of Appendix \ref{sec:extraction_of_PSD}). We therefore conclude that the $f^{-1}$ spectrum arises from charge noise and SEC provided by the micromagnet stray field\cite{Yoneda2017}. Indeed, from a Ramsey measurement performed in the condition optimized for the large quality factor (large-SEC regime), we extract the low-frequency ($f<100\,\text{Hz}$) noise in line with the $f^{-1}$ dependence (Fig.~\ref{fig:spectroscopy}b). Similar $f^{-1}$ noise spectrum over seven decades of frequency has been observed in an isotopically purified $^{28}$Si device\cite{Yoneda2017}.
Approximating $S(f)=A^{2}/f$, however, we find $A\sim 0.6\,\text{MHz}$ being two orders of magnitude larger than $A\sim 1.6\,\text{kHz}$ found in the $^{28}$Si device. It is also an order of magnitude larger than $A\sim 0.1\,\text{MHz}$ observed in a GaAs device without micromagnet\cite{Malinowski2017a}. The difference can be partly attributed to large SEC in the present device, as $S(f)$ is reduced by two orders of magnitude by decreasing SEC (see Fig.~\ref{fig:spectroscopy}b), at least at low frequencies. SEC also depends on the orbital energy splitting determined by confinement potential and effective mass.
However, the influence of other factors on $A$, such as material properties and experimental setups, requires further investigations.

To summarize, we demonstrate $24$-fold enhancement of the dephasing time $T_2^*$ of a GaAs single electron spin qubit. The enhancement relies on suppressing the nuclear spin noise by feedback and is limited by classical control electronics that can be improved further. The feedback also boosts the qubit overall performance: We reach the quality factor of $Q=85\pm 8$ and the $\pi$-gate fidelity of $99.04\pm 0.23\,\%$. We find that, despite our device being a GaAs quantum dot, the ultimate fidelity with the feedback is not limited by nuclear-spin noise. The culprit is the $1/f$ charge noise leaking into the qubit through the micromagnet field gradient at megahertz frequencies.


\appendix
\section{Additional details of the feedback protocol}
Our feedback control protocol is implemented on a Xilinx ZedBoard device equipped with a coupled central processing unit and programmable logic. The device is interfaced with an AD-FMCOMMS2-EBZ peripheral board from Analog Devices which provides integrated RF demodulators, 12-bit digital-to-analog converters (DACs) with a sampling rate up to $122.88\,\text{M}$ samples per second, and local oscillators (LOs) operating at up to $6\,\text{GHz}$. The demodulators and DACs are used to digitize the RF charge-sensing signal for single-shot spin readouts, and one of the LOs generates the driving microwave at $f_\text{MW}$ for the EDSR. This system enables rapid switching of $f_\text{MW}$ conditioned on the spin measurement outcomes without overhead for the data transfer between different equipments. The off-resonant microwave burst is generated from a discrete signal generator and fed to the same signal path via a microwave switch. The whole setup is described in the Supplemental Material.

For the realtime feedback control, we first take $150$ single-shot data points of a Ramsey oscillation with varied intervals $t_\text{R}=2,4,\cdots 300\,\text{ns}$ in the ``probe'' step. Each measurement sequence shown in Fig.~\ref{fig:ramsey}b takes $T_\text{R}=31.71\,\mu\text{s}$, giving $T_\text{p}=150\,T_\text{R}\approx 4.8\,\text{ms}$ for one probe step. We update the value of $\fest$ based on $\delta f$ obtained from the Bayesian estimation\cite{Shulman2014,Delbecq2015} and adjust the microwave frequency to $f_\text{MW}=\fest+\Delta$ for the ``target'' experiment. We wait for $T_\text{w}$ before starting the target step, and we find that $T_\text{w}$ needs to be longer than $2\,\text{ms}$ to stabilize the microwave output. The time $T_\text{t}$ spent in the target step depends on the type of the experiment and it is $T_\text{t} = T_\text{p}$ in the case of the measurement shown in Fig.~\ref{fig:feedback}. Thus, we define the feedback latency between the probe and target steps as $\Delta t = T_\text{p}/2 + T_\text{w} + T_\text{t}/2$, which can be controlled by changing $T_\text{w}$.

\section{Decoherence of spin qubits}
Here we summarize our model of spin (qubit) decoherence in free and driven evolutions using the formulation in Ref.~\onlinecite{Ithier2005}, which is independent of the microscopic origin of the noise.

\subsection{Decay in free evolution}
The spin coherence during free evolution is characterized by the decay envelope of Ramsey oscillations, also called a free induction decay. The dynamics of a spin is described by two processes, the longitudinal relaxation (depolarization) and the pure dephasing. The longitudinal relaxation leads to an exponential decay with the rate $\Gamma_1 = T_1^{-1}$. The Fermi's golden rule gives $\Gamma_1 = \pi^2\ST (f_\text{qubit})$, where $\ST (f_\text{qubit})$ is the power spectral density (PSD) of the transverse noise at the qubit resonance frequency $f_\text{qubit}$. On the other hand, the pure dephasing is caused by the longitudinal noise (fluctuation of the qubit frequency $\delta f$), with the PSD given by $\SL (f)=\int_{-\infty}^\infty \left< \delta f(t) \delta f(t+\tau) \right> \exp(-i 2\pi f \tau) d\tau$. If the noise is short correlated (white noise), it leads to an exponential decay with the rate $\Gamma_\varphi = 2\pi^2 \SL (0)$. In this case, the transverse relaxation (dephasing) of the free evolution is given by an exponential envelope with the rate $\Gamma_2 = T_2^{-1}=\frac{1}{2}\Gamma_1 + \Gamma_\varphi$.

The pure dephasing of a spin qubit is often dominated by a longer correlated noise, with its spectral density increasing at lower frequencies. The decay envelope is then generally non-exponential, which we here express as the decoherence function $W(t)$. For Gaussian noise, $W(t)$ is expressed as
\begin{equation}
W(t) = \exp\left( -\frac{t^2}{2}(2\pi)^2 \int_{-\infty}^\infty df\,\SL(f)\,\mathrm{sinc}^2 (\pi f t) \right),
\label{eq:W}
\end{equation}
with $\mathrm{sinc}\, x = \sin x / x$.
We divide $W(t)$ into two parts, $W(t)=W^\text{static}(t) W^\text{high}(t)$, where $W^\text{static}$ describes the contribution from the quasi-static noise ($|f|<1/t$) and $W^\text{high}$ describes the higher-frequency contribution ($|f|>1/t$).
Combined with the longitudinal relaxation, the decay envelope of the free evolution is given by
\begin{equation}
E^\text{free}(t) = W^\text{static}(t)W^\text{high}(t)\exp(-\Gamma_1 t/2).
\label{eq:ramseydecay}
\end{equation}
In the pure dephasing of the free evolution, the effect of $W^\text{high}$ can be neglected ($W^\text{high}(t)=1$).
The quasi-static part is evaluated using $\mathrm{sinc}(\pi f t) \sim 1$ for $|f|\ll 1/t$ as
\begin{equation}
W^\text{static}(t) \approx \exp\left( -\frac{t^2}{2} (2\pi\sigma)^2 \right),
\end{equation}
where $\sigma^2 = \lim_{T\rightarrow\infty}\frac{1}{T}\int_0^T |\delta f(t)|^2\,dt = \int_{-\infty}^\infty \SL(f)\,df$ is the variance of the qubit frequency detuning $\delta f$. This corresponds to the Gaussian decay with $T_2^* = 1/(\pi\sqrt{2}\sigma)$ usually obtained in the quasi-static approximation, where $\delta f$ is considered constant during each qubit evolution but Gaussian-distributed in an experimental run. This approximation is valid in our case, where $\sigma$ and $T_2^*$ are dominated by the quasi-static noise due to the Overhauser field fluctuation with $\SL(f)\propto f^{-\beta}$ and $\beta\sim 2$.
The feedback control with the latency $\Delta t$ described in the main text sets a low-frequency cutoff of $\SL(f)$ at $f_\text{c}\sim 1/\Delta t$, thereby reducing $\sigma^2$ and enhancing $T_2^*$. On the other hand, the feedback would be less effective for $\beta\leq 1$, where noise at higher frequencies contributes more to $T_2^*$.

\subsection{Decay in driven evolution}
Now we turn to the spin coherence during driven evolution, considering a measurement of Rabi oscillations.
The spin dynamics is described in the frame rotating with the driving frequency $f_\text{MW}$. The spin rotates around the vector sum of the driving field $f_\text{rabi}$ along the $x$ axis and the microwave frequency detuning $\Delta_\text{q} = f_\text{MW} - f_\text{qubit}$ along the $z$ axis. The total field is tilted by an angle $\eta = \arctan (f_\text{rabi}/\Delta_\text{q})$ from the $z$ axis and its length is $f_\text{R}=\sqrt{f_\text{rabi}^2 + \Delta_\text{q}^2}$. Decoherence in driven evolution is considered by separating the quasi-static ($|f|<1/t$) part and the higher-frequency ($|f|>1/t$) part of the noise. The Rabi decay envelope is given by
\begin{equation}
E^\text{rabi}(t) = W^\text{static}_\text{rabi}(t)W^\text{high}_\text{rabi}(t)\exp(-\tilde{\Gamma}_2 t),
\end{equation}
with $\tilde{\Gamma}_2$ the transverse relaxation rate in the rotating frame.

The contribution of the higher-frequency noise is calculated by mapping $\SL(f)$ and $\ST(f)$ to the components transverse and longitudinal with respect to the total field in the rotating frame. Assuming $\eta$ is constant ($\Delta_\text{q}=\Delta$ and $f_\text{qubit}=\fest$), the longitudinal relaxation rate in the rotating frame is
\begin{equation}
\tilde{\Gamma}_1 = \sin^2 \eta\, \Gamma_\nu + \frac{1+\cos^2 \eta}{2} \Gamma_1,
\end{equation}
with $\Gamma_\nu = 2\pi^2 \SL(f_\text{rabi})$. Similarly, $\ST(f_\text{qubit})$ contributes to the pure dephasing via $\tilde{\Gamma}_\varphi = \frac{1}{2}\Gamma_1 \sin^2 \eta$, leading to the transverse relaxation rate
\begin{equation}
\tilde{\Gamma}_2 = \frac{1}{2}\tilde{\Gamma}_1 + \tilde{\Gamma}_\varphi = \frac{3-\cos^2 \eta}{4}\Gamma_1 + \frac{1}{2}\Gamma_\nu \sin^2 \eta .
\end{equation}
The longitudinal noise $\SL(f)$ also contributes to $W^\text{high}_\text{rabi}(t)$, which is obtained by replacing $\SL(f)$ in Eq.~\eqref{eq:W} with $\SL(f)\cos^2 \eta$ and setting a low-frequency cutoff to $1/t$.

The contribution of the quasi-static noise is calculated by averaging the Rabi oscillations with the Gaussian-distributed noise of $\Delta_\text{q}$ around $\Delta$, $\Delta_\text{q} - \Delta = \fest - f_\text{qubit} = -\delta f$. At zero detuning $\Delta = 0$, this gives a power-law decay\cite{Cucchietti2005}
\begin{equation}
W_\text{rabi}^\text{static}(t) = \left[ 1 + \left(2\pi\frac{\sigma^2 t}{f_\text{rabi}}\right)^2 \right]^{-1/4}.
\end{equation}
In addition, this averaging leads to an initial phase shift\cite{Koppens2007} in the Rabi oscillation, which, however, is negligibly small for $f_\text{rabi}\gg \sigma^2 t$ relevant here.
Since $W_\text{rabi}^\text{high}(t)=1$ for $\Delta = 0$ and $|\delta f|\ll f_\text{rabi}$, the decay envelope at zero detuning is simplified to
\begin{equation}
E_0^\text{rabi}(t) = W_\text{rabi}^\text{static}(t) \exp\left[- \left(\frac{3}{4}\Gamma_1 + \frac{1}{2}\Gamma_\nu\right)t\right].
\label{eq:rabidecay}
\end{equation}
For $\SL(f_\text{rabi}) \gg \ST (f_\text{qubit})$, it is $\Gamma_\nu \propto \SL(f_\text{rabi})$ that dominates the exponential part.

\begin{figure}
\includegraphics[width=0.48\textwidth]{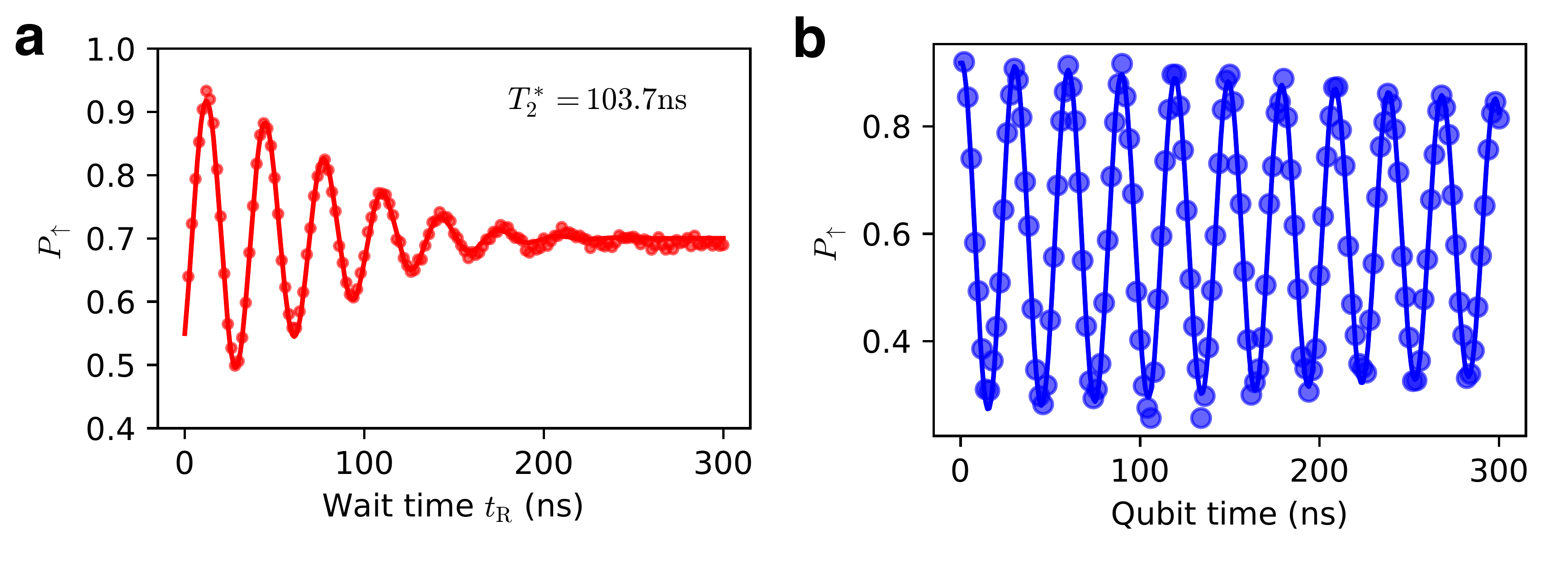}
\caption{Coherent oscillations with a large SEC. These data are taken in a gate-bias condition similar to those of Figs.~\ref{fig:control} and \ref{fig:spectroscopy}, where SEC is increased.
(a) Ramsey oscillation. Spin-up probability $P_\uparrow$ is measured as a function of $t_\text{R}$ with the feedback on ($\Delta = 30\,\text{MHz}$). The solid curve is a fit to the Gaussian-decaying oscillation with $T_2^* = 103.7\,\text{ns}$.
(b) Rabi oscillation for $Q=85$. The solid curve is a fit to the exponentially decaying oscillation with $f_\text{rabi} = 33.64 \pm 0.01\,\text{MHz}$ and $T_2^\text{rabi} = 1.26 \pm 0.12\,\mu\text{s}$.
\label{fig:ramsey_SEC}}
\end{figure}

\section{Extraction of the noise power density $S(f)$}
\label{sec:extraction_of_PSD}
To increase the qubit control fidelity by increasing $f_\text{rabi}$, a larger spin-electric coupling (SEC) is favorable. In the present device, we can tune the SEC by gate-controlled inter-dot energy detuning which could change either the electron confinement or the local slanting magnetic field\cite{Tokura:2006ir}.
However, a larger SEC also enhances the susceptibility of the spin qubit to charge noise, leading to a larger PSD of the qubit frequency noise.
After increasing the SEC for Figs.~\ref{fig:control} and \ref{fig:spectroscopy}, we notice that the dephasing time is indeed reduced to $T_2^* \sim 100\,\text{ns}$ as shown in Fig.~\ref{fig:ramsey_SEC}a.
This suggests that the dephasing time observed in free evolution (a Ramsey experiment) is dominated by charge noise in this regime.

To extract $\SL(f)$ in Fig.~5 by fitting the Rabi decay envelopes with Eq.~\eqref{eq:rabidecay}, we calculate $\sigma^2$ from the Ramsey decay envelope.
If $\SL(f)=A^2/f$ holds in the whole frequency range of interest, $f_\text{c}<|f|<1/t$, and $\Gamma_1$ is neglected, the decay envelope in Eq.~\eqref{eq:ramseydecay} reduces to
\begin{equation}
E^\text{free}(t)\approx W^\text{static}(t) = \exp\left[-t^2 (2\pi A)^2 \ln \frac{1}{f_\text{c}t} \right]
\end{equation}
with $1/T_2^* = 2\pi A \sqrt{\ln\frac{1}{f_\text{c}t}}$.
Using $T_2^*=103.7\,\text{ns}$, $f_\text{c}=0.5\,\text{kHz}$ and $t=100\,\text{ns}$, we obtain $A=0.5\,\text{MHz}$. In the Rabi decay, on the other hand, the high-frequency cutoff is determined by $t\sim T_2^\text{rabi}\sim 2\,\mu\text{s}$. The variance relevant to $W_\text{rabi}^\text{static}$ is therefore estimated to be $\sigma^2 = 2 A^2 \ln\frac{1}{f_\text{c}t} = 3.3\,\text{MHz}^2$.

The expression of $T_2^*$ in the above also explains why the dephasing time shown in Fig.~\ref{fig:ramsey_SEC}a remains so short with the feedback control. The $T_2^*$ value is improved only logarithmically by increasing the low-frequency cutoff $f_\text{c}$.
Nevertheless, the feedback control is essential to suppress quasi-static noise especially during driven evolution. As shown by Eq.~\eqref{eq:rabidecay}, significant portion of $\SL(f)$ in $W_\text{rabi}^\text{high}(t)$ is decoupled from the Rabi decay by maintaining $\Delta = 0$ using the feedback control, so that one can reach the $\Gamma_\nu$-limited regime.

\begin{figure}
\includegraphics[width=0.48\textwidth]{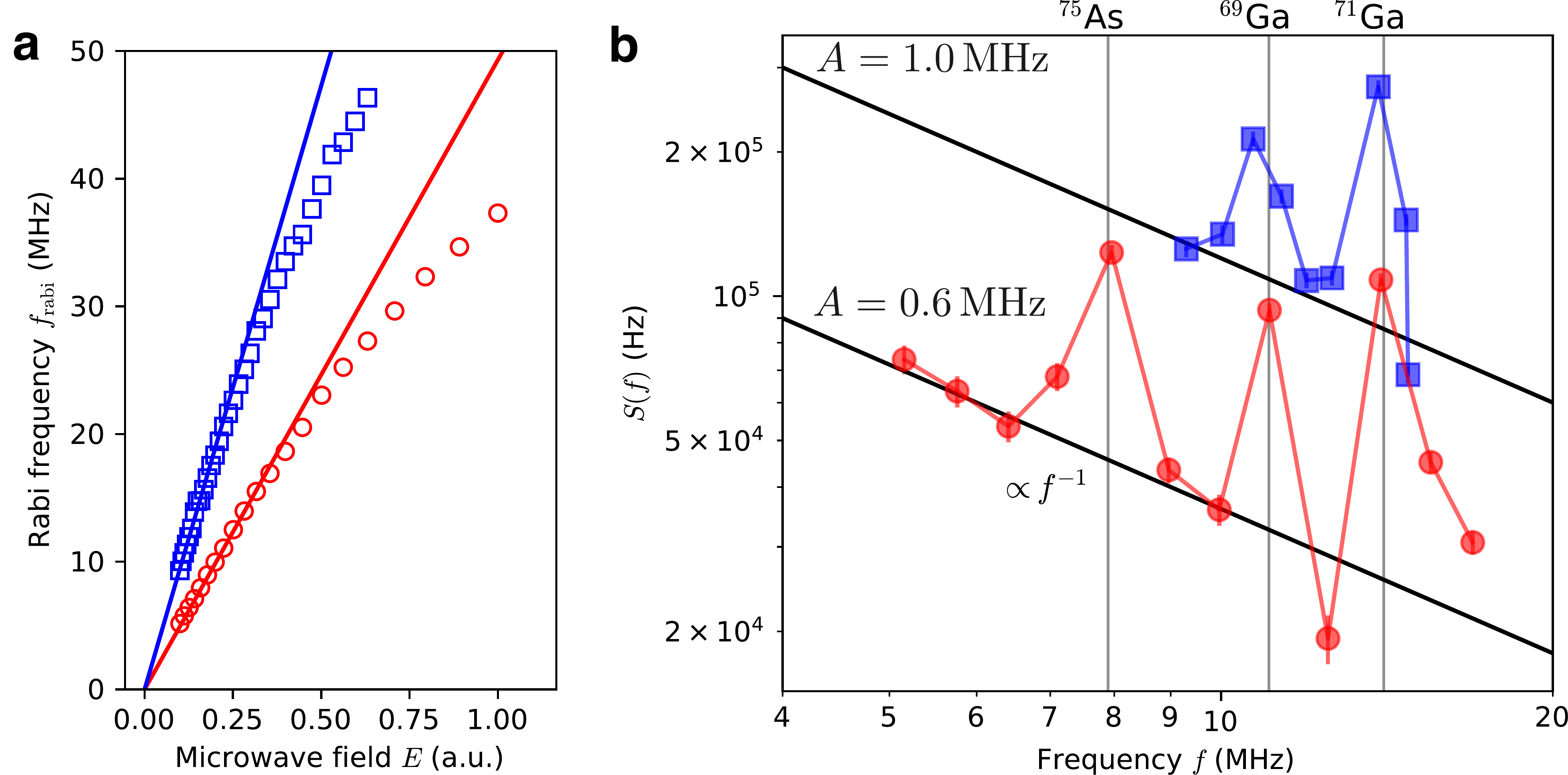}
\caption{Microwave amplitude dependence of the Rabi frequency (a) and Rabi noise spectroscopy (b) with different SECs.
Red circles are taken in the ``large SEC'' condition with $A=0.6\,\text{MHz}$ as shown in Fig.~\ref{fig:spectroscopy}. Blue squares are taken in a condition where SEC is increased to $A=1.0\,\text{MHz}$ and the $Q$ factor is slightly lower.
\label{fig:sec_comparison}}
\end{figure}

To examine the validity of the analysis, we have analyzed another set of data taken at an even larger SEC. Figure~\ref{fig:sec_comparison} shows the comparison of the extracted PSD and the one plotted in Fig.~\ref{fig:spectroscopy}a. Due to the increased (transverse) SEC, the Rabi frequency is $1.9$ times larger at the same driving microwave field. The high-frequency noise PSD is also increased as shown in Fig.~\ref{fig:sec_comparison}b. The value of $A$ is found to increase by a factor of $1.7$ in agreement with the increase of the longitudinal SEC. This consistency confirms that the main source of the Rabi decay is the high-frequency charge noise coupled to the spin qubit.

\acknowledgments
We thank S.~D.~Bartlett for fruitful discussions.
We thank RIKEN CEMS Emergent Matter Science Research Support Team and Microwave Research Group at Caltech for technical assistance.
Part of this work was financially supported by
CREST, JST (JPMJCR15N2, JPMJCR1675),
the ImPACT Program of Council for Science, Technology and Innovation (Cabinet Office, Government of Japan),
JSPS KAKENHI Grants No. 26220710, No. 18H01819, No. 19K14640 and No. 16K05411,
RIKEN Incentive Research Projects,
The Murata Science Foundation Research Grant,
and MEXT Quantum Leap Flagship Program (MEXT Q-LEAP) Grant Number JPMXS0118069228.
T.O. acknowledges support from
JSPS KAKENHI Grants No. 16H00817 and No. 17H05187,
PRESTO (JPMJPR16N3), JST,
Yazaki Memorial Foundation for Science and Technology Research Grant,
Advanced Technology Institute Research Grant,
Izumi Science and Technology Foundation Research Grant,
TEPCO Memorial Foundation Research Grant,
The Thermal \& Electric Energy Technology Foundation Research Grant,
The Telecommunications Advancement Foundation Research Grant,
Futaba Electronics Memorial Foundation Research Grant,
and MST Foundation Research Grant.
A.D.W. and A.L. greatfully acknowledge support from BMBF-Q.Link.X 16KIS0867, DFG-TRR160, and DFH/UFA CDFA-05-06.

%


\end{document}